\documentclass[11pt,a4paper]{article}
\pdfoutput=1

\usepackage{graphicx}
\usepackage{amsmath}
\usepackage{amssymb}
\usepackage{geometry}
\usepackage{authblk}
\usepackage{hyperref}
\usepackage{nameref}

\def\sv{\left\langle\sigma v\right\rangle}
\def\({\left(}
\def\){\right)}

\title{Fermi-LAT kills dark matter interpretations of AMS-02 data. Or not?\\}

\author[1]{Konstantin Belotsky}
\author[1]{Ruslan Budaev}
\author[1,2]{Alexander Kirillov}
\author[1,3]{Maxim Laletin}
\affil[1]{National Research Nuclear University MEPhI, 115409 Kashirskoe shosse 31, Moscow, Russia}
\affil[2]{Yaroslavl State P.G. Demidov University, 150000 Sovietskaya st., 14, Yaroslavl, Russia}
\affil[3]{Space sciences, Technologies and Astrophysics Research (STAR) Institute, Universit\'{e} de Li\`{e}ge, B\^{a}t B5A, Sart Tilman, 4000 Li\`{e}ge, Belgium}
\date{}

\begin{document}

\maketitle

\begin{abstract}
	
A number of papers attempt to explain the positron anomaly in cosmic rays, observed by PAMELA and AMS-02, in terms of dark matter (DM) decays or annihilations. However, the recent progress in cosmic gamma-ray studies challenges these attempts. Indeed, as we show, any rational DM model explaining the positron anomaly abundantly produces final state radiation and Inverse Compton gamma rays, which inevitably leads to a contradiction with Fermi-LAT isotropic diffuse gamma-ray background measurements. Furthermore, the Fermi-LAT observation of Milky Way dwarf satellites, supposed to be rich in DM, revealed no significant signal in gamma rays.

We propose a generic approach in which 
the major contribution to cosmic rays comes from the dark matter disc
and prove that the tension between the DM origin of the positron anomaly and the cosmic gamma-ray observations can be relieved.
We consider both a simple model, in which DM decay/annihilate into charged leptons, and a model-independent minimal case of particle production, and we estimate the optimal thickness of DM disk. Possible mechanisms of  formation and its properties are briefly discussed.  

\end{abstract}

\section{Introduction}

 One of the reasons to assume that dark matter (DM) is more than just an obscure non-luminous substance revealing itself only by virtue of its gravitational attraction is connected with the interpretation of the high-energy cosmic positron excess, discovered by PAMELA \cite{Adriani:2008zr} and accurately confirmed by AMS-02 \cite{Aguilar:2013qda}. 
 The zest of this observation is that no known mechanism of cosmic-ray production or acceleration can provide the increase of positron fraction at high energies (comparing to electrons) and thus an additional source of energetic positrons is needed. 
 While some researchers consider astrophysical objects like pulsars or supernova remnants (e.g. see \cite{Hooper:2008kg,DiMauro:2014iia}) as the origin of this phenomenon, the others point out to DM being involved in it (e.g. see \cite{Bai:2009ka,Chen:2015cqa}). Both types of explanations have their pros and cons and both can be constrained by current and future experiments. As for DM, these constraints mostly come from direct-detection experiments, CMB observations, cosmic antiproton flux measurements and, last but not least, cosmic gamma-ray surveys, including the studies of dwarf galaxies and isotropic diffuse gamma-ray background (IGRB) (see \cite{2015arXiv151102031C} for a review). The latest Fermi Large Area Telescope (LAT) results show no significant gamma-ray signal from Milky Way satellite galaxies, which are considered to be the most dark-matter-dominated astrophysical objects \cite{Ackermann:2013yva}, and the measured spectrum of IGRB \cite{Ackermann:2014usa} also seems to be very limitative (especially taking into account their recent conclusion that up to 86\% of extragalactic gamma-ray background beyond 50~GeV can be due to unresolved sources \cite{DiMauro:2016cbj}). But why is it so important to emphasize this point, whereas the other constraints look no less serious? In fact, it is no big deal to construct a dark matter model which explains the positron anomaly and fulfils the requirements of underground experiments (e.g. see \cite{Fox:2008kb,Dev:2013hka}) and CMB observations \footnote{CMB constraints can be evaded if the cross section is suppressed at low energies or early times \cite{2015arXiv151102031C}.}, and, though difficult, it is possible to restrain antiproton production within the model. But it is practically impossible to have a self-consistent DM model, which would produce enough positrons to explain the data and won't end up with an overabundance of gamma rays, accompanying DM decays or annihilations into a pair of charged leptons or emerging during the propagation of these charged leptons in the interstellar media. 
 This problem was somehow revealed in a series of papers (see e.g. \cite{Cirelli:2012ut,Ando:2015qda,Liu:2016ngs}).

However, taking a closer look at the problem one may notice the following. As is known, while high-energy positrons are born throughout the dark matter halo only those produced in the $\sim 3$~kpc vicinity can contribute significantly to the cosmic ray (CR) spectrum and result in the high-energy positron excess. This is due to the peculiar features of charged particles propagation in the Galactic magnetic media. 
Gamma rays on the other hand do not interact with magnetic fields and come to us directly from the whole DM halo. This fact is the root of the aforementioned problem. In other words, if Nature wanted some particles to be the source of positron anomaly she could sprinkle them locally in sufficient amount and since there are no unnecessary particles in the halo (which would give no positrons, but only gamma) the average flux of gamma rays over the same solid angle would be drastically suppressed. This means that DM can be divided in (at least) two components: a ``faint'' or ``passive'' component, which makes up the major part of the DM halo and does not (significantly) contribute to observable CR, and an ``active'' one, which is subdominant, explains most of the positron anomaly and for some reason is concentrated nearby. 
In fact, this idea doesn't imply that active and passive DM components should be of different physical origin, as we discuss in the \nameref{discus} section.
	
The most natural way to introduce such a local source distribution seems to be to put active DM in the Galactic disk or to suppose that this type of DM forms a disk-like structure itself. Surprisingly enough, the same concept of a DM disk was proposed earlier to explain various phenomena, such as global flaring of the Galactic hydrogen disk \cite{Kalberla:2007sr} or the planar structure of DM rich dwarf satellites around Andromeda \cite{Randall:2014kta}. Moreover, as simulations show \cite{Read:2008fh,Purcell:2009yp,Kuhlen:2013tra}, the formation of a dark matter disk can be the result of satellite accretion into the Milky Way stellar disk (though, according to these simulations, such a disk hardly contributes to the local DM density). Observational data \cite{Bidin:2010rj,Ruchti:2015bja,Bienayme:2014kva,Xia:2015agz,2016arXiv160401407K} are far from making unambiguous conclusions on the existence of dark disk (DD). 

Despite the fact that the effects of DD on CR were studied in the past \cite{Cholis:2010px,Evoli:2011id}, we provide a novel argument for the existence of DM disk: it is the vital element needed to avoid gamma-ray constraints on the DM positron anomaly explanation. Actually, a disk is not the only possible form of DM substructure providing the necessary effect~--- there might be some dense local DM clump or bubble \cite{Hektor:2013yga}, though this possibility stays beyond the scope of this paper.

In this article, following the ideas of our previous investigations \cite{Alekseev:2016off,Alekseev:2016axt} we present the results of a cumulative statistical analysis of both the AMS-02 cosmic positron fraction data and the Fermi-LAT data on IGRB under the hypothesis of DM, decaying or annihilating into $e^+e^-,\mu^+\mu^-$ and $\tau^+\tau^-$. We show explicitly how the introduction of an active DM disk reduces the gamma-positron ratio at high energies and significantly improves the goodness-of-fit. 
As an addition, we consider %a model-independent
 the least model-dependent minimal case of DM positron production, tuned to explain AMS-02 positron fraction data, and demonstrate that an isotropic distribution of active DM component throughout the whole dark halo is most likely ruled out by IGRB measurements. We conclude with a brief discussion of DD properties and some particular DM models providing the mechanisms of DD formation.

\section{Dark halo case}

First, we are going to consider %the common case, where 
that active DM particles are distributed ubiquitously in the dark halo (we refer to it as the ``halo case'' further in the text), %Following the convention of numerous studies, 
for which we use Navarro-Frenk-White (NFW) density profile \cite{Navarro:1996gj}. One should note, though, that the choice of DM distribution does not significantly affect our results since most of the reasonable DM profiles are quite similar in the $\sim 3$~kpc vicinity of the Sun (see e.g. Fig.~1 in \cite{Cirelli:2010xx}), from where the major contribution into positron flux comes from, or at high Galactic latitudes ($\left|b\right| > 20^{\circ}$), where Fermi-LAT measures the IGRB. 

The current analysis is performed under a simple model, in which only $e^+e^-$, $\mu^+\mu^-$ and $\tau^+\tau^-$ decay/annihilation channels are allowed. 
Initial (injection) spectra of $e^{\pm}$ and prompt gamma rays from DM decay/annihilation are simulated through Pythia 8 \cite{Sjostrand:2007gs}, 
and the effects of positron propagation in the Galaxy are calculated using the GALPROP code \cite{GalpropCode}.
We adopt the propagation parameters that provide the best fit to AMS-02 data on cosmic protons and to the $B/C$ ratio \cite{Jin:2014ica}.
The fluxes of positrons (and electrons) resulting from different decay/annihilation channels %($e^+e^-$, $\mu^+\mu^-$, $\tau^+\tau^-$)
are calculated independently and then summed with the corresponding branching ratios. %($Br_{e^+e^-}$, $Br_{\mu^+\mu^-}$, $Br_{\tau^+\tau^-}=1-Br_{e^+e^-}-Br_{\mu^+\mu^-}$). 
Background fluxes of cosmic positrons and electrons are taken from \cite{Ibarra:2009dr}. The effects of Solar modulation are accounted for using the charge sign-dependent force-field model \cite{2009...ICRCGast} %with different potentials for $e^-$ and $e^+$: $\phi_{e^-}=700~\text{MV}$, $\phi_{e^+}=800~\text{MV}$ 
(these effects are, however, negligible in the data we are considering, i.e. above 30~GeV).

Gamma rays associated with DM decays/annihilations can be divided into two groups: prompt gamma radiation, which comes straight from the interaction vertex, and gamma rays induced by the propagation of $e^{\pm}$ in the Galaxy. The latter, related to such processes as inverse Compton scattering (ICS) and Bremsstrahlung, are calculated using GALPROP.
The averaged flux of prompt gamma rays %(coming from the Galactic regions, which fall into the field of sight of Fermi-LAT, when it observes IGRB)
 is given by
% \begin{widetext}
\begin{equation}
%	\Phi_{\rm prompt}\left(E\right)=\frac{1}{4\pi\Delta \Omega}\int\limits_{\Delta \Omega} d\cos(b)\;dl \int\limits_{0}^{S(b,l)} ds \; j(s,b,l) \sum_{i}^{} {\rm Br}_i  \: f^i_{\gamma}\left(E\right).
	\Phi_{\rm p}\left(E\right)=\frac{1}{4\pi\Delta \Omega}\int\limits_{\Delta \Omega} d\Omega \int\limits_{0}^{S} d\vec{s} \; j(\vec{s}) \sum_{i}^{} {\rm Br}_i  \: f^i_{\gamma}\left(E\right).
\end{equation}
 %\end{widetext}
Here $f^i_{\gamma}\left(E\right)$ denotes the differential energy spectrum of prompt photons, emitted in the i-th channel with branching ratio ${\rm Br}_i$, 
%of the corresponding channel, 
$S$ stands for the approximate distance %at which the DM distribution `
to the DM halo boundary, % is located
$\Delta \Omega$ is the observable solid angle ($20^{\circ} \leq \left|b \right| \leq 90^{\circ}$ and $0^{\circ} < l \leq 360^{\circ}$), which falls into the field of sight of Fermi-LAT when it observes IGRB%within the range of Galactic coordinates 
. In our notations $j(\vec{s})$ stands for the emissivity factor, which we define as follows:

\begin{equation}
j(\vec{s}) = 
 \begin{cases}
   \cfrac{\sv}{\xi M^2}\, \rho^2(\vec{s}) &\text{(annihilation)},\\
   \cfrac{1}{M\tau}\, \rho(\vec{s}) &\text{(decay),}
 \end{cases}
\end{equation}
%Total gamma flux includes contributions 
where $\sv$ stands for the velocity-averaged annihilation cross section %\footnote{Here we consider DM particles being Dirac fermions; for Majorana DM emissivity is larger by a factor of 2. }
$\xi %stands for the numerical factor, which is equal to 
	= 2$ for Majorana DM particles and $\xi = 4$ for Dirac DM particles, $M$ denotes the mass of the active DM particle and $\tau$ is its lifetime in case of decays. 
For the sake of convenience we treat the local emissivity factor $j_\text{loc} = j(0)$ as a free parameter, which equally defines the production rate of $e^{\pm}$ and gamma both for decays and annihilations.
Though prompt radiation also comes from extragalactic sources, in this work we do not include the extragalactic contribution, since its estimation relies significantly on the velocity and density distributions of active DM in the outer Universe, making it model-dependent. %As we do not attach ourselves to a specific model of interacting DM, but only want to perform a generic comparison of cases (i) and (ii), the extragalactic contribution is not included. 
However, we would like to emphasize that our predictions are at most 2 times lower than they should be (since the Galactic and extragalactic contributions are comparable) and furthermore that even then the overabundance of gamma rays from the halo is evident. %We also discuss a few points related to the extragalactic gamma-ray contribution below.

To find the parameter values, which provide the best possible fit to cosmic positron data while taking into account the IGRB constraints, the following %easily comprehensible
approach %, based on the concept of Pearson's chi-square test, 
is used: we introduce a cumulative test-statistic, which depends on the model parameters (i.e. the mass of the particle $M$, the branching ratios ${\rm Br}_i$, the local emissivity $j_\text{loc}$) and on the experimental data on the cosmic positron fraction (AMS-02 \cite{Accardo:2014lma}) and on the IGRB (Fermi-LAT \cite{Ackermann:2014usa}): 

\newcommand{\fr}[1]{ F^{(\text{#1})} }
\newcommand{\phigamma}[1]{ \Phi^{(\text{#1})} }
% \begin{widetext}

\begin{equation}
	\newcommand{\fri}[1]{ F^{(\text{#1})}_i }
	\newcommand{\phigammaj}[1]{ \Phi^{(\text{#1})}_j }
	\chi^2 = \sum_{i=1}^{k}\left(\dfrac{ F\left(E_i,\vec{p}\right)-\fri{exp} }{ \sigma_i}\right)^2 + \sum_{j=1}^{m} \theta\left(\Phi\left(E_j,\vec{p}\right)-  \phigammaj{exp}\right) \left(\dfrac{  \Phi\left(E_j,\vec{p}\right)-\phigammaj{exp} }{ \sigma_j }\right)^2.
	\label{chi2}
\end{equation}
% \end{widetext}
%is minimized.

Here indices $i$ and $j$ enumerate AMS-02 and Fermi-LAT experimental data points respectively, while $k$ and $m$ denote the numbers of experimental points under analysis, $\vec{p}$ denotes the set of model parameters listed above,  $F(E_i,\vec{p})$ and $\Phi(E_j,\vec{p})$ are the predicted cosmic positron fraction and gamma-ray flux, %(calculated at certain experimental energy dots), %\footnote{Gamma-ray flux was calculated by averaging over solid angle...}, 
$\fr{exp}_i$ and $\phigamma{exp}_j$ are the corresponding experimental values with the respective errors $\sigma_i$ and $\sigma_j$, and $\theta$ stands for the Heaviside step function. Since the IGRB data gives an upper limit on the gamma-ray flux from DM annihilations/decays, the theoretically predicted curve should not fit the experimental data points, but lie beneath them. Thus the second term in Eq. \ref{chi2} contributes to the test statistic only when the predicted gamma-ray flux exceeds the data for at least one point. This means that even if some set of parameters provides a perfect explanation of cosmic positron data the considered method can, nevertheless, rule it out due to the predicted overabundance of gamma rays. On the other hand, an attempt to reduce the amount of gamma would likely spoil the fit of positron fraction data and, thus, there should be an optimal case, which can be found by minimizing $\chi^2$. Besides, the test statistic itself provides a valuable information about the ``goodness-of-fit'' %\footnote{ Though a test statistic can be transformed into $p$-value, which is commonly thought to reflect the probability of the hypothesis to be true from the experimental point of view, here we do not consider it due to a risk of misinterpretation.}
%We do not transform the obtained $\chi^2$ values into P-values, since our estimate Eq.\eqref{chi2} does not correspond in strict sense to conventional $\chi^2$, however definitely reflects agreement level.  
~--- if $\chi^2/n < 3$, where $n = k+m-{\rm dim}(\vec{p})$ denotes the number of degrees of freedom (in our analysis the total amount of data points minus $3$ parameters gives $n = 59$),  %(with ${\rm dim}(\vec{p})$ being the number of free parameters)
we assume that the fit quality is satisfactory. % and the larger $\chi^2/n$ exceeds $1$, the worse is the level of agreement. 
%Total amount of datapoints taken minus $3$ (number of free parameters) give us number of degrees of freedom $59$.} 

%\begin{equation}
%	\newcommand{\fri}[1]{ F^{(\text{#1})}_i }
%	\newcommand{\phigammaj}[1]{ \Phi^{(\text{#1})}_j }
%	\chi^2 = \sum_{i=1}^{k}\left(\dfrac{ F\left(E_i,\vec{p}\right)-\fri{exp} }{ \sigma_i}\right)^2 + \sum_{j=1}^{m} \theta\left(\Phi\left(E_j,\vec{p}\right)-  \phigammaj{exp}\right) \left(\dfrac{  \Phi\left(E_j,\vec{p}\right)-\phigammaj{exp} }{ \sigma_j }\right)^2.
%	\label{chi2}
%\end{equation}
% \end{widetext}
%is minimized.

\begin{figure}[t!]
	\begin{minipage}[h]{0.5\linewidth}
		\center{\includegraphics[width=1\textwidth]{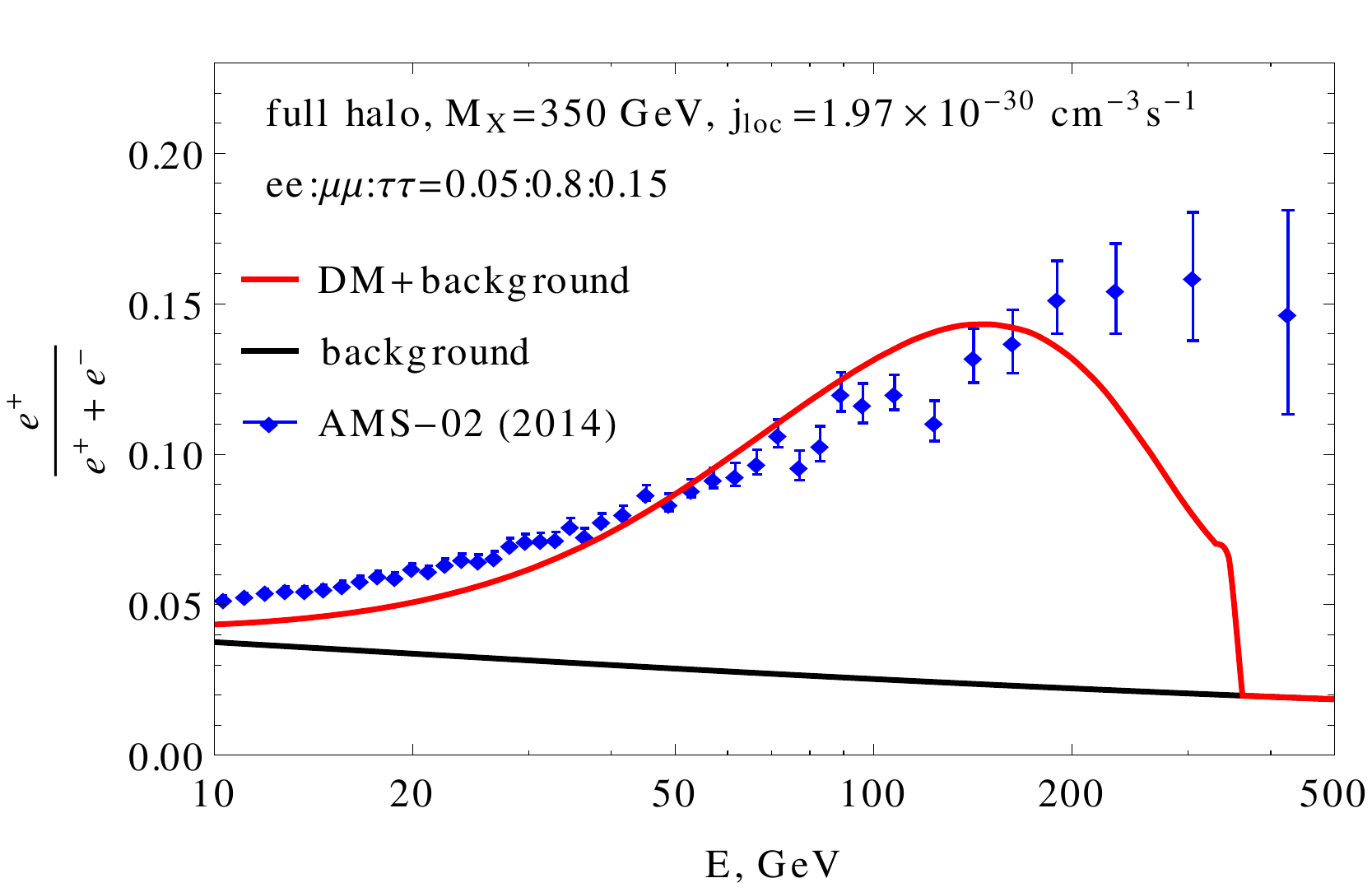} a)\\}
	\end{minipage}	
	\hfill
	\begin{minipage}[h]{0.48\linewidth}
		\center{\includegraphics[width=1\textwidth]{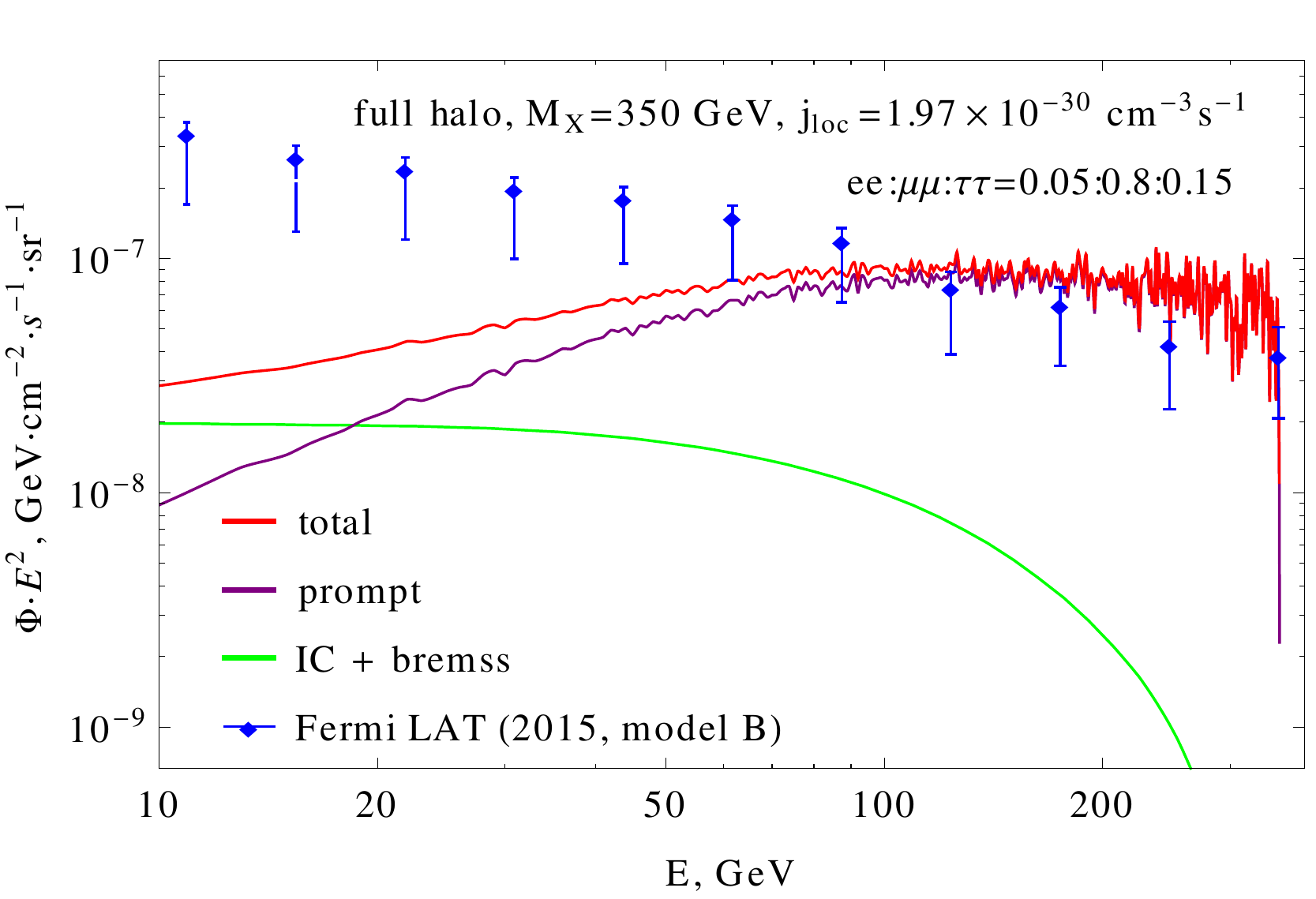} b)\\}
	\end{minipage}
	\vfill
		\begin{minipage}[h]{0.5\linewidth}
			\center{\includegraphics[width=1\textwidth]{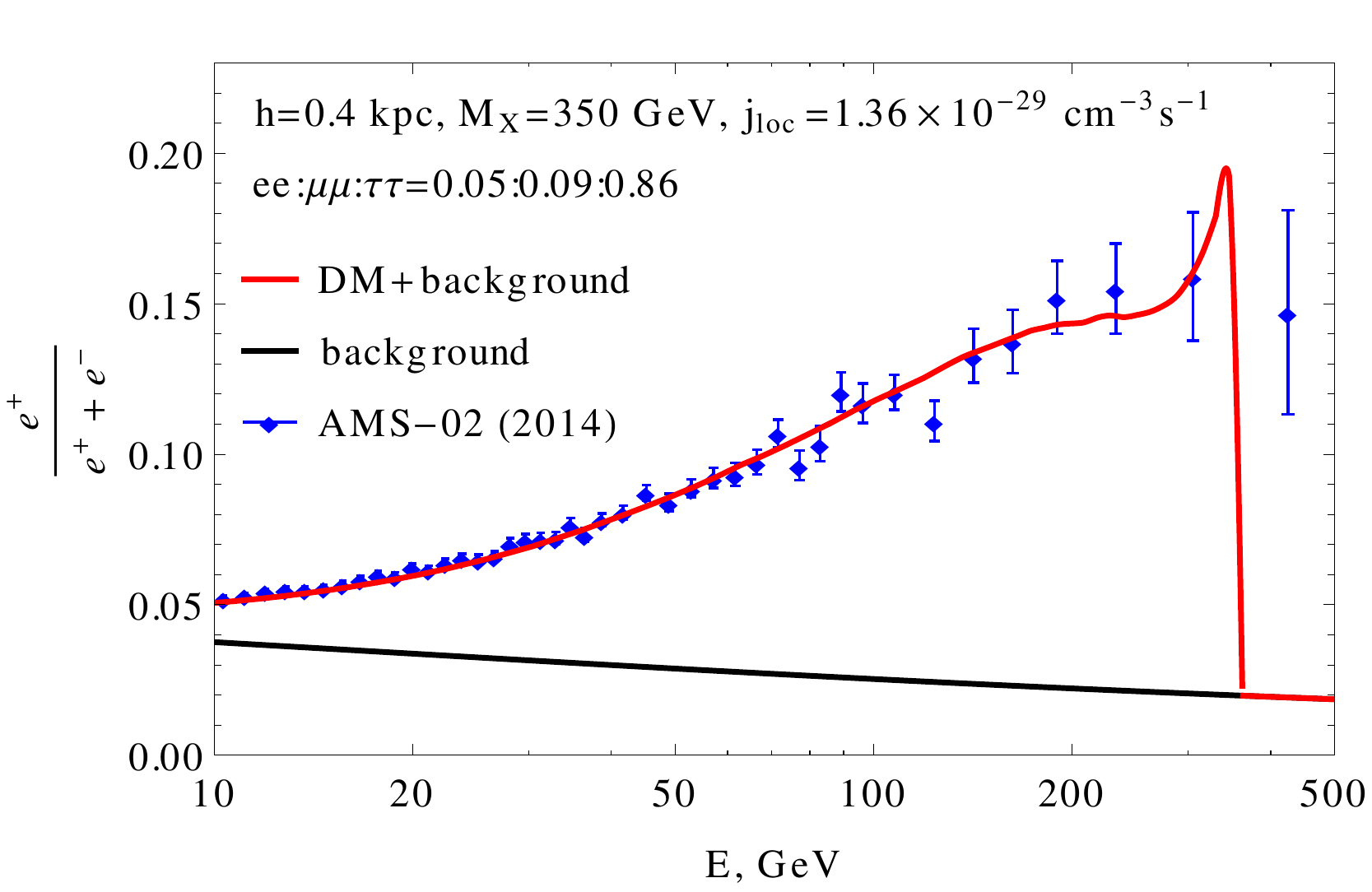} c)\\}
		\end{minipage}	
		\hfill
		\begin{minipage}[h]{0.48\linewidth}
			\center{\includegraphics[width=1\textwidth]{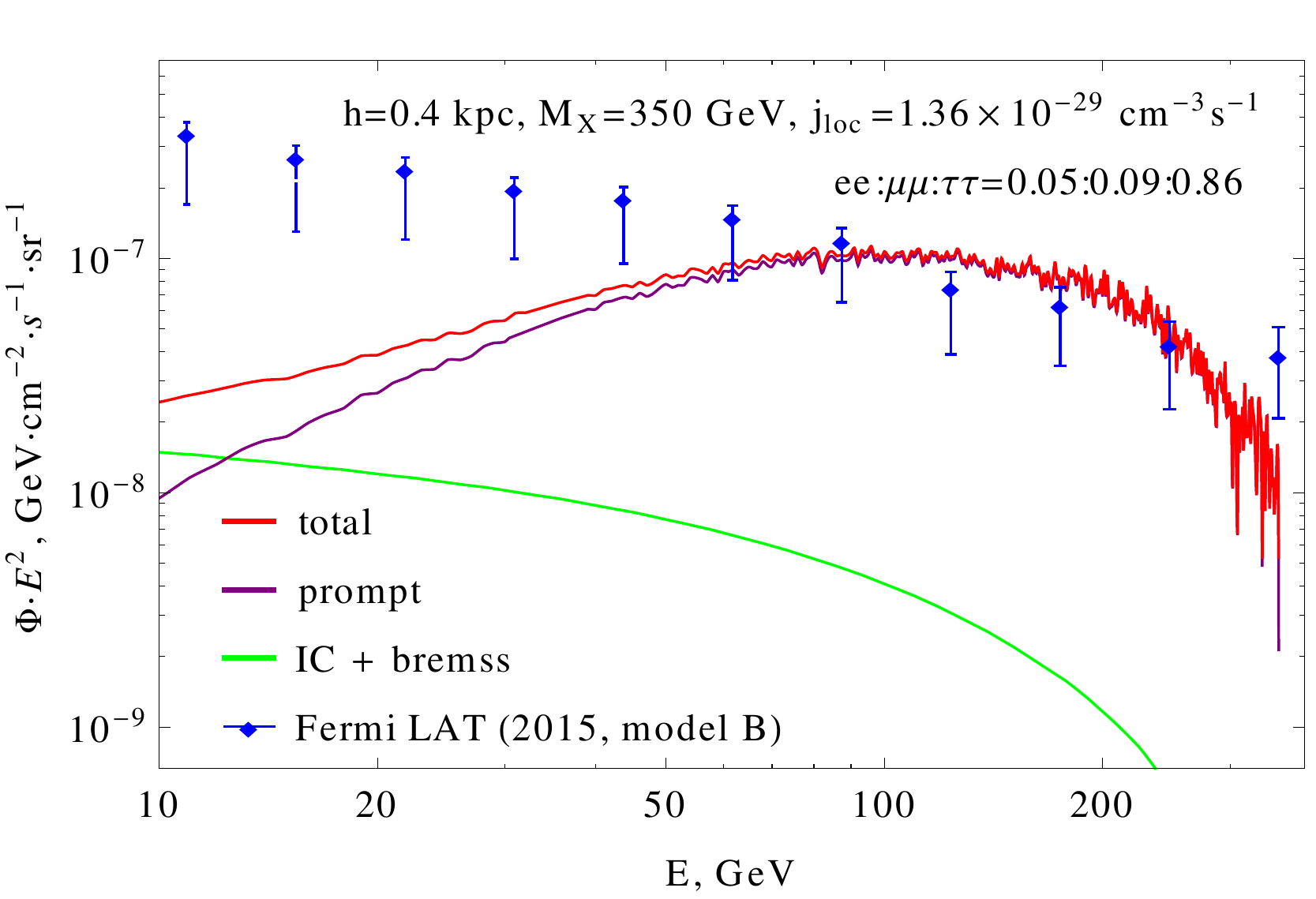} d)\\}
		\end{minipage}	
		
	\caption{The best possible fit of the positron fraction (left) measured by AMS-02 \cite{Accardo:2014lma} and the isotropic gamma-ray flux (right), measured by Fermi-LAT (IGRB, model B) \cite{Ackermann:2014usa}. Top: dark halo case. Bottom: dark disk case. Fluxes of gamma-rays are multiplied by $E^2$. The values of model parameters are listed \textit{on top} of each plot (the given value of mass corresponds to the case of annihilating DM).}
	\label{spectra}
\end{figure}

Using the method described above, we numerically calculate the least possible value of $\chi^2$. We get $\chi_{\rm min}^2/n \approx 5$ for annihilations and $\chi_{\rm min}^2/n \approx 8$ for decays, which means that the halo case is excluded at $> 99\%$ C.L. The best \textit{possible} fit of positron fraction and isotropic gamma-ray flux are shown in Fig. \ref{spectra} (a,b).  Though the obtained positron fraction does not look as the ``best fit'' at all, it has a simple statistical explanation: the method we use prefers to degrade the fit of the high-energy positron fraction to fulfil the hard gamma-ray constraint, since (due to the large experimental errors in AMS-02 data above $\sim 200$~GeV) it results in lower values of $\chi^2/n$. 

We would like to stress that, although, the model we use in our analysis is trivial, it is rather demonstrative. Fitting with pure $e^+e^-$, $\mu^+\mu^-$, $\tau^+\tau^-$ modes result in much worse $\chi^2/n$ values \footnote{Note, that in \cite{Liu:2016ngs} a %softer criterion was used, 
different approach was used, which didn't account for the quality of the positron data fit and allowed a $\mu^+\mu^-$-mode.}. We do not consider any other modes (such as $W^+W^-, b\bar{b} \text{ or } \gamma\gamma$), since they produce redundant antiprotons and gamma.

\section{Dark disk case}

Let us now consider the case where the active DM component is only concentrated in the disk (which we refer to as the ``disk case''). %For the same reason the case of DM disk (meaning that ``active'' DM component is only concentrated in the disk) 
This case can be simulated by simply cutting the NFW profile at $z=\pm h$, where $z$ denotes the longitudinal axis in a cylindrical coordinate system and $2h$ is the assumed value of the DD thickness. Thus, an additional parameter $h$ is introduced. 
Though such clipped NFW profile seems %generally inappropriate
unphysical %for the dark disk, %(mostly because of an unexpected central cusp),
it allows to compare directly the cases under analysis and demonstrate a pure (one-parameter) clipping effect. %From the physical point of view this corresponds to the ``pressing'' of ``active'' DM component into a disk. 
One should however be more careful when one chooses the density profile of the DM disk, especially if the region under consideration includes the Galactic center. %It seems that, in general, an exponential profile is appropriate (e.g. Eq. 3 in \cite{Read:2008fh}).

\begin{figure}[h!]
%	\begin{minipage}[h]{0.5\linewidth}
%		\center{\includegraphics[width=1\textwidth]{Posi_450_halo.pdf} a)\\}
%	\end{minipage}	
%	\hfill
%	\begin{minipage}[h]{0.48\linewidth}
%		\center{\includegraphics[width=1\textwidth]{Gamma_450_halo.pdf} b)\\}
%	\end{minipage}
%	\vfill
		\begin{minipage}[h]{0.5\linewidth}
			\center{\includegraphics[width=1\textwidth]{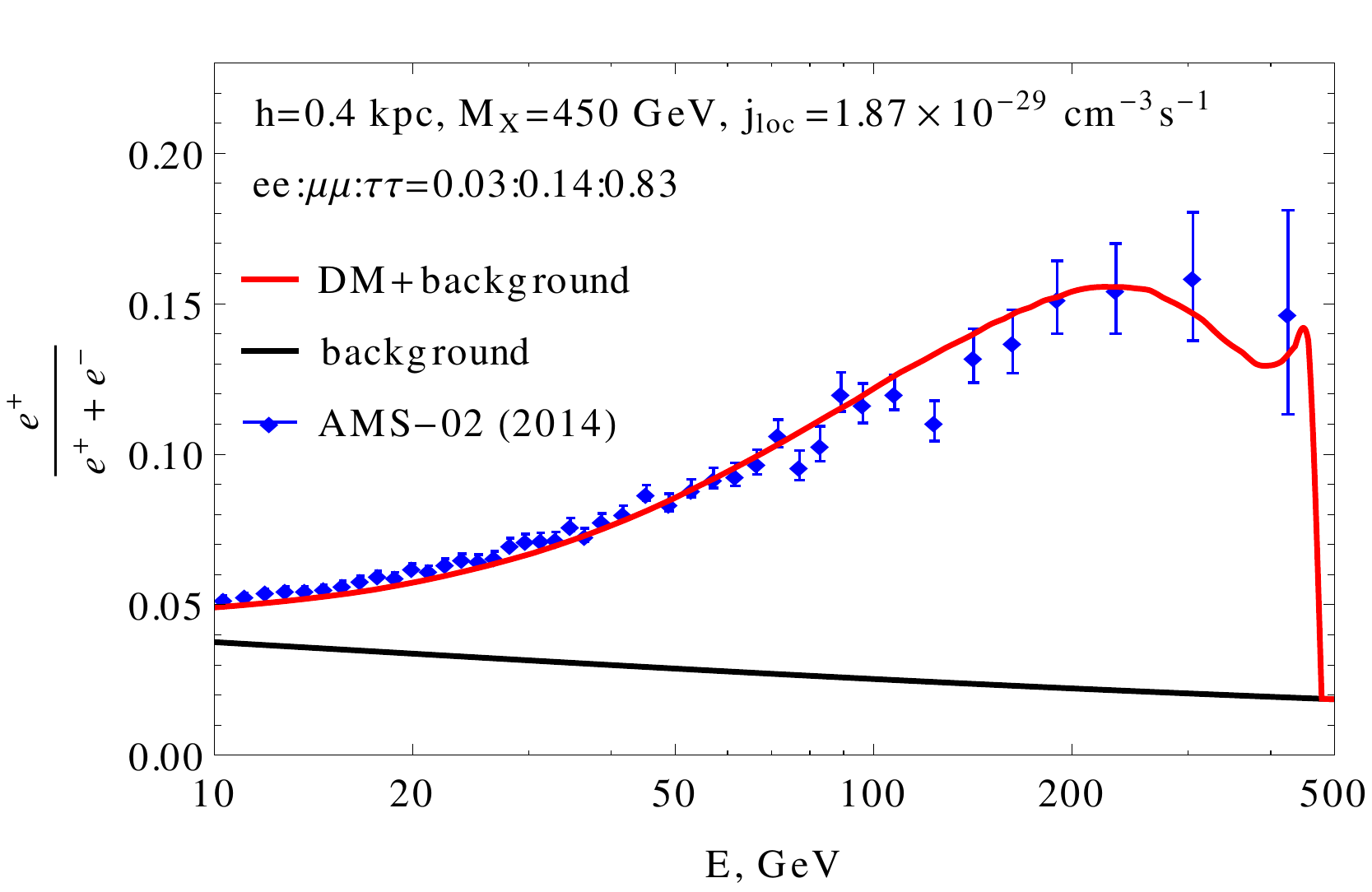} a)\\}
		\end{minipage}	
		\hfill
		\begin{minipage}[h]{0.48\linewidth}
			\center{\includegraphics[width=1\textwidth]{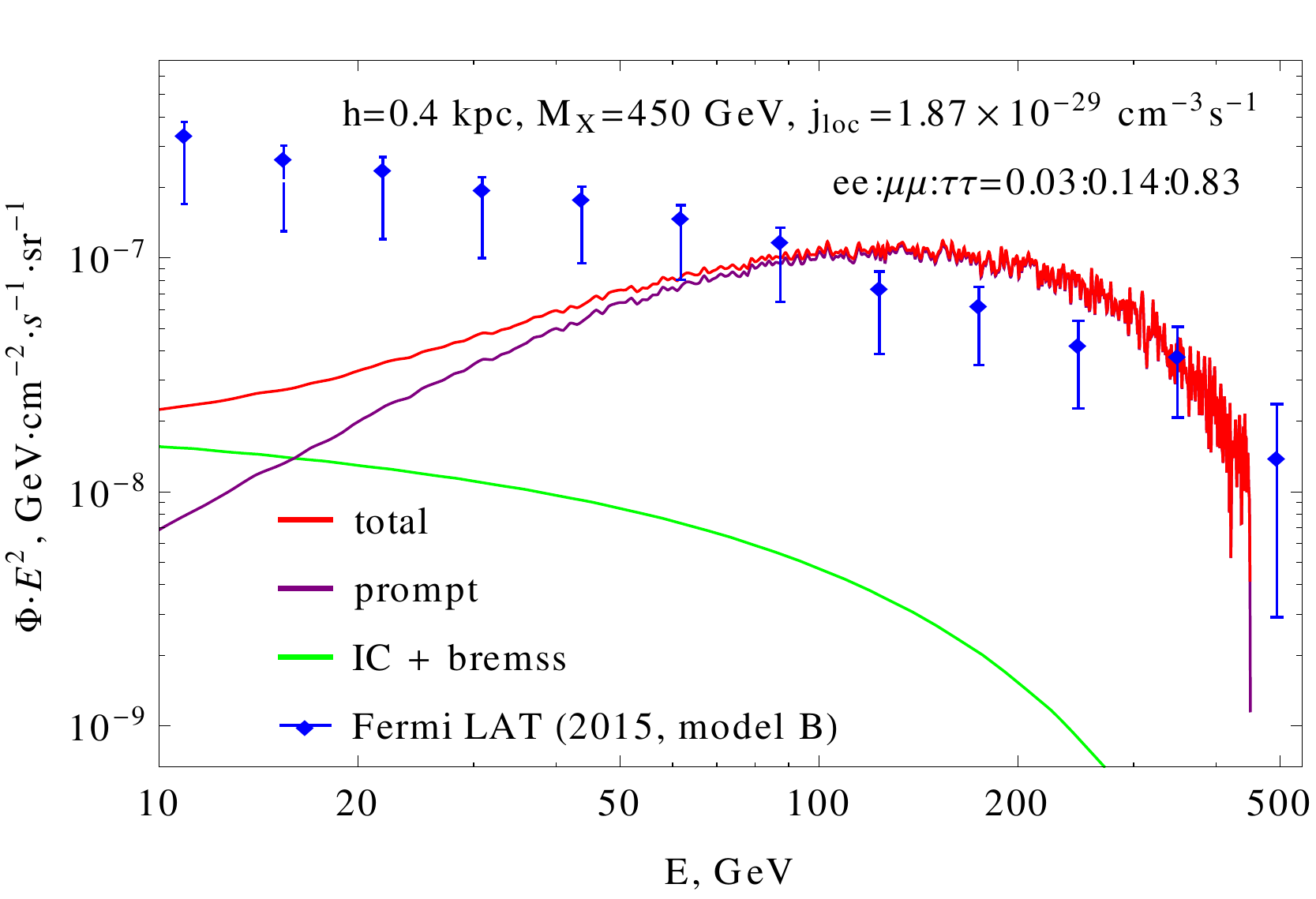} b)\\}
		\end{minipage}	
		
	\caption{Spectra of positron fraction and gamma-rays as in Fig.~\ref{spectra} (bottom line) for $M_X = 450$ GeV.}
	\label{spectra_450}
\end{figure}

\begin{figure}[p]

%	\begin{minipage}[h]{0.6\linewidth}
%	\center{\includegraphics[width=1\textwidth]{chiSq_emutau_a.pdf}}
%	\end{minipage}
%	\vfill
%	\begin{minipage}[h]{0.6\linewidth}
%	\center{\includegraphics[width=1\textwidth]{chiSq_emutau_d.pdf}}
%	\end{minipage}

	\center{\includegraphics[width=0.7\textwidth]{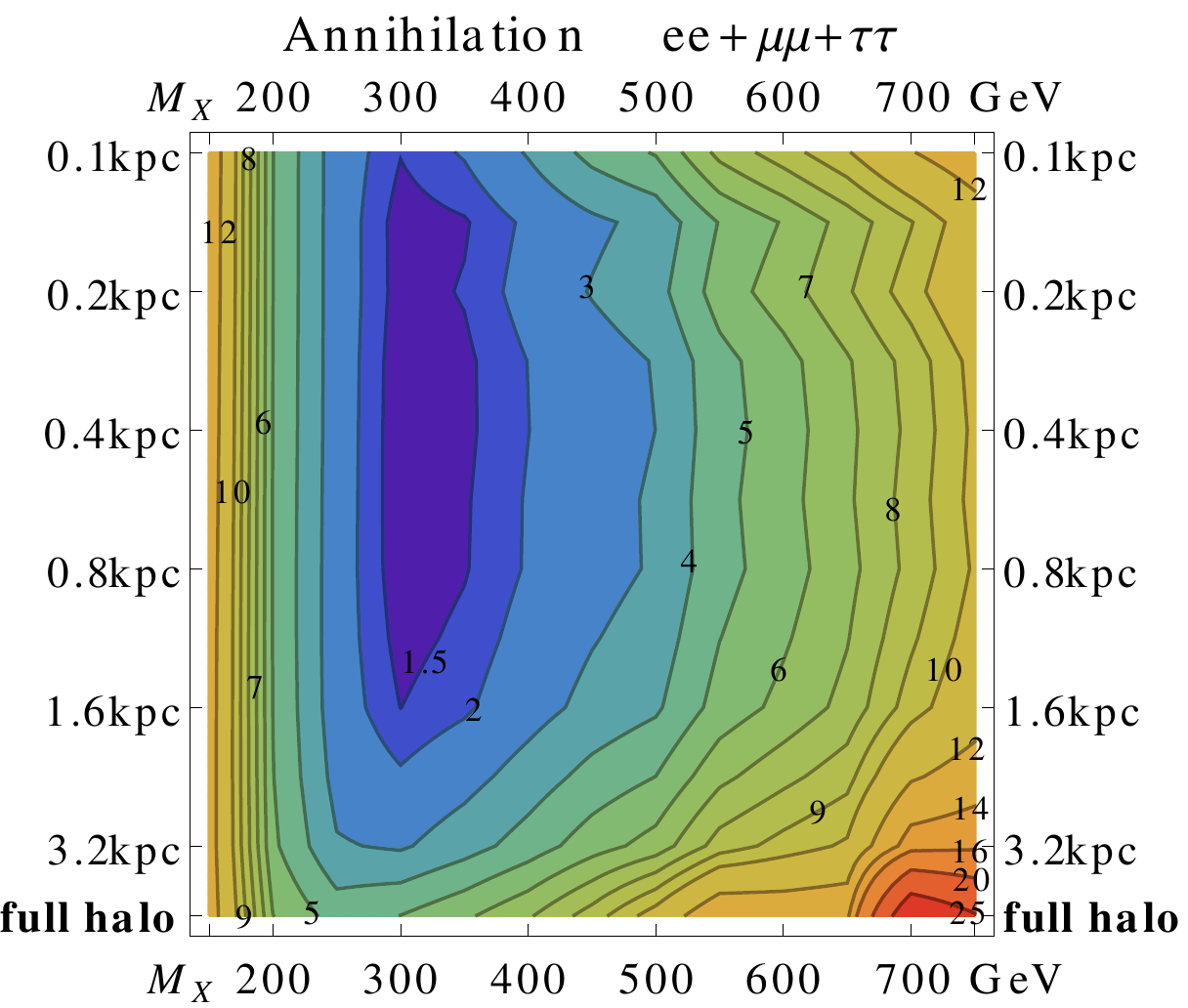}}

	\vfill

	\center{\includegraphics[width=0.7\textwidth]{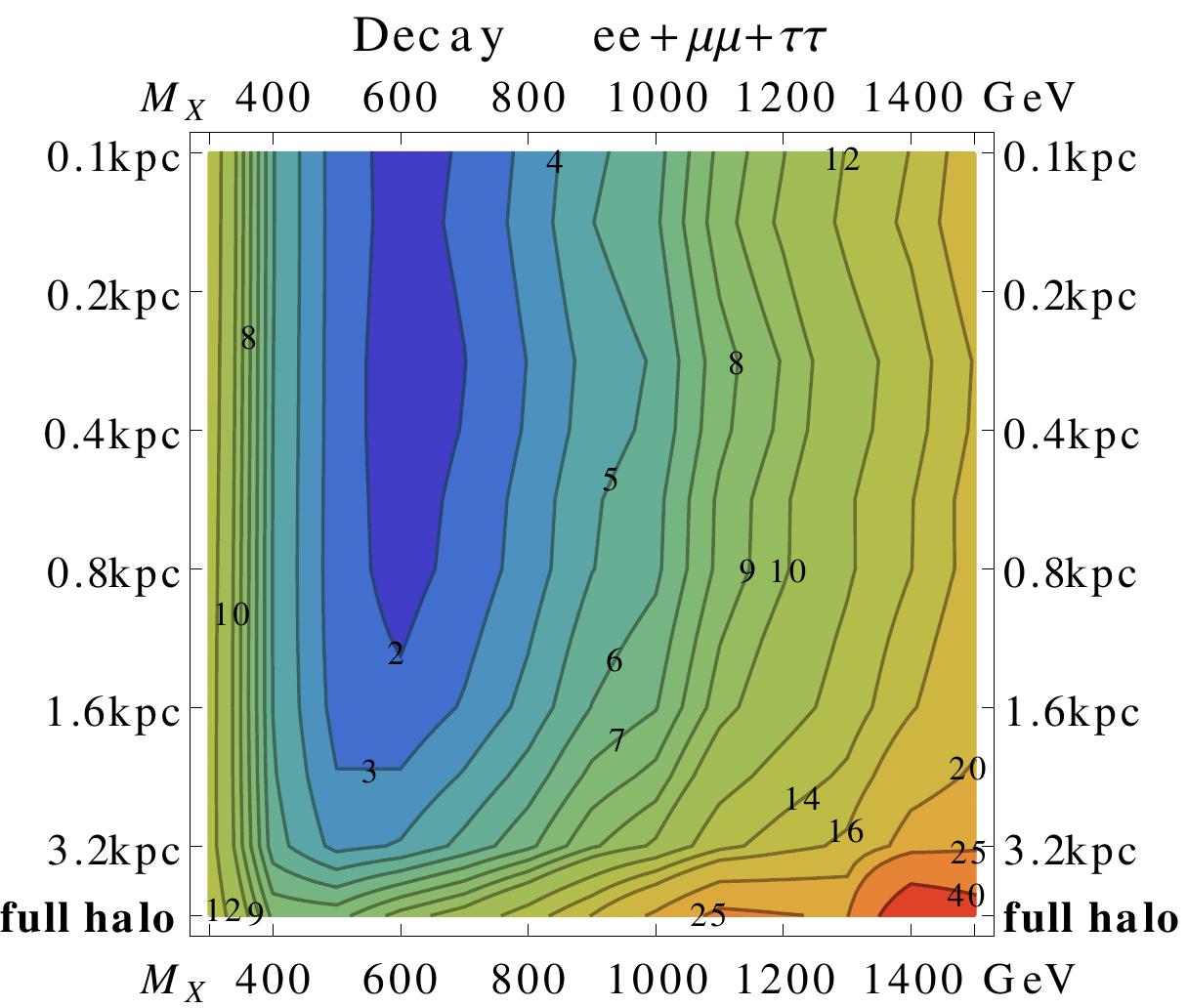}}
	
	\vfill

	\caption{$\chi^2/n$ dependence of DM mass $M$ and DD half-height $h$ for annihilation (top) and decay (bottom) into $e^+e^-, \, \mu^+\mu^- \text{and } \tau^+\tau^-$. The color gradient flow from red to blue corresponds to the decrease of $\chi^2/n$. The numbers on the plot contours denote the respective $\chi^2/n$ values. In the area below $h = 3.2$~kpc the $\chi^2/n$ values, corresponding to the halo case, are shown.
	}
	\label{chi2fig}
\end{figure}

The contour plots in Fig. \ref{chi2fig} show the $\chi^2/n$ dependence on DM particle mass $M$ and on the DD half-height $h$ %for DM annihilating/decaying into $\tau^+\tau^-$ only (a,b) and 
for DM annihilating/decaying into $e^+e^-, \, \mu^+\mu^- \text{and } \tau^+\tau^-$ with optimal branching ratios. For each fixed value of $M$ and $h$ we obtain the lowest possible values of $\chi^2/n$, using the method described above. For comparison, on the lowest edge of the plots we give $\chi^2/n$ values corresponding to the halo case. As one can clearly see, the disk case results in much lower values of $\chi^2/n$. The range of best-fit parameter values corresponds to $M \approx 300$~GeV  and $h \approx 0.1 - 2$~kpc for annihilations, and $M \approx 600$~GeV  and $h \approx 0.1 - 1$~kpc for decays. Spectra of positron fraction and gamma rays corresponding to the least value of $\chi^2/n$ are shown in Fig.\ref{spectra} (c,d). As we have mentioned above the minimal $\chi^2/n$ case fails to fit the AMS-02 data point $> 400$ GeV, but one can improve the positron data fit at the cost of $\chi^2/n$ increase %For comparison, we also show the same curves for the halo case (c,d).
%\textbf{Fig.~\ref{spectra_450} shows a better fit of positron fraction data in the dark disk case, though the value of $\chi^2/n$ is greater than the minimal one.}
(see Fig. \ref{spectra_450}).

The thinner DD reduces the amount of produced high-energy positrons, while the larger one leads to the steady increase of diffuse gamma-ray flux. 

Other properties of active DM in the DD depend on extra assumptions. For example, assuming that local DD density is close to its upper observational limit, i.e. $\sim 0.3\div 0.4$ GeV/cm$^3$ \cite{Bienayme:2014kva,Xia:2015agz} (comparable to the halo DM contribution), one gets $\sv\sim (2\div 3)\times 10^{-23}$ cm$^3/$s (for Majorana DM) in the best fit case. Thus, the annihilation cross-section is $500\div 1000$ times boosted comparing to the commonly assumed thermally averaged cross-section for WIMPs. %It is remarkable that 
The resulting active DM fraction of the cosmological DM density would be $(1\div 2)\times 10^{-3}$, if active DM have a WIMP-like thermal evolution (though separate from passive DM) and if there is no annihilation enhancement in the Galaxy. %(or the enhancement is saturated)
%It is remarkable, this fraction value agrees 
%which 
It is remarkable, that this result is less than one order of magnitude close to the estimated DD contribution to the mass of the Galaxy, assuming the exponential density profile of DD (see e.g.~\cite{Read:2008fh}). Furthermore, the considered realization of active DM model successfully evades the constraints on annihilating DM imposed by Planck data \cite{Ade:2015xua}.

One may notice that, according to Fig. \ref{spectra} d) and \ref{spectra_450} b), the ICS gamma-ray contribution lies well beneath the Fermi-LAT experimental limits. However, the authors of \cite{Cholis:2010px} obtained a very different result for the dark disk case. Given the model parameters considered in that paper we made an attempt to reproduce their calculations, but our result for ICS gamma-ray flux is smaller by almost two orders of magnitude, while the signal in positrons is the same. We performed a cross-check of the method we use and successfully reproduced the results of \cite{Cirelli:2009vg}, where the ICS gamma-ray flux was calculated using analytical approximations. %Following the approach described in the latter article we estimated the ICS gamma-ray flux for the dark disk case, which is also consistent with our numerical calculations.

\section{The minimal case}

One may argue that a more complex model of DM would be able to produce enough high-energy positrons to fit the AMS-02 data without the corresponding excess in gamma rays. %For example, a decay/annihilation into a pair of some new intermediate particles, which consequently decay into $e^+e^-$ or $\mu^+\mu^-$, might mimic the $\tau$-mode positron spectra, while yielding less gamma. 
However, we are going to consider what we shall refer to as the minimal case of DM particle production and show that even in this case the problem with IGRB still remains. 

 %We used the following two-parametric function with a cut-off at $1 \TeV$:
Despite the fact that in general the gamma-ray spectrum depends on the details of the model, % a simple %fact
the following argument can be used to impose the \textit{least model-dependent} constraint. %no matter how a %charged particle
%\textbf{positron} is produced, it emits FSR.
The key requirement to the models we consider is to reproduce the data on charged CR, which means that they all should yield relatively the same spectrum of positrons. Being a charged particle, positron gives its contribution to the gamma-ray spectrum of the process. Hence, this is the only contribution, which would be present in any model (besides the possible contributions from intermediate charged particles, hadronic decays, etc.) and which we call the minimal gamma-ray spectrum. Essentially, without the fine-tuning of the intermediate particle masses this contribution depends only on the energy spectrum of the outcoming positrons.
Thus, the minimal gamma-ray spectrum %accompanying positron production defined by Eq. \ref{minspec},
can be %calculated
estimated as %follows
a convolution of the positron spectrum $f_e$ and the FSR spectrum, emitted by the ``single" positron with energy $E_0$ (see e.g. Eq.~47 in \cite{Mardon:2009rc} with $x_0 = E/E_0$ and $\epsilon_0 = m_e/E_0$):
%which allows, after GALPROP calculation, to fit AMS data on positrons very well \cite{...}. Then injection photon spectrum is given by

% \begin{widetext}
\begin{equation}
\label{mingamma}
f_{\gamma}(E) = \frac{\alpha}{\pi E} \int\limits_E^{\text{1 TeV}} \left( 1 + \left( 1 - \frac{E}{E_0} \right)^2 \right) \left( \ln\left[ \left( \frac{2E_0}{m_e} \right)^2 \left( 1 - \frac{E}{E_0} \right) \right] -1 \right)\, f_e(E_0) \; dE_0, %\\
\end{equation}
% \end{widetext}
with $\alpha$ the fine-structure constant and $m_e$ the electron mass. %We use the energy spectrum of FSR corresponding to the case of $e^{\pm}$ pair production (see e.g. \cite{Essig:2009jx}). %.GALPROP gives also 
This formula gives a good approximation at high energies, where the Fermi-LAT constrains IGRB the most.

\begin{figure}[h!]
	\begin{minipage}[h]{0.5\linewidth}
		\center{\includegraphics[width=1\textwidth]{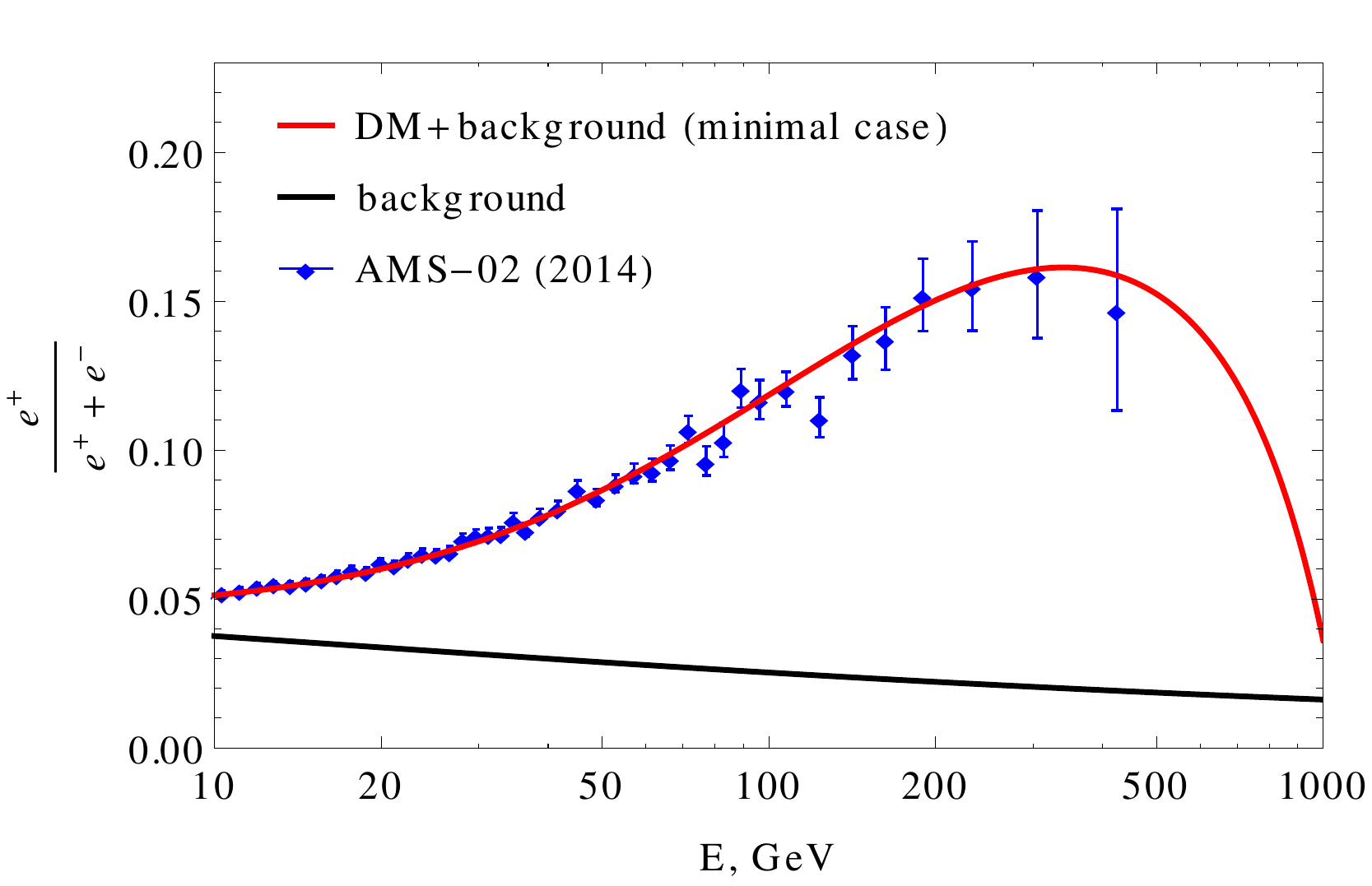} a)}
	\end{minipage}
	\hfill
	\begin{minipage}[h]{0.47\linewidth}
		\center{\includegraphics[width=1\textwidth]{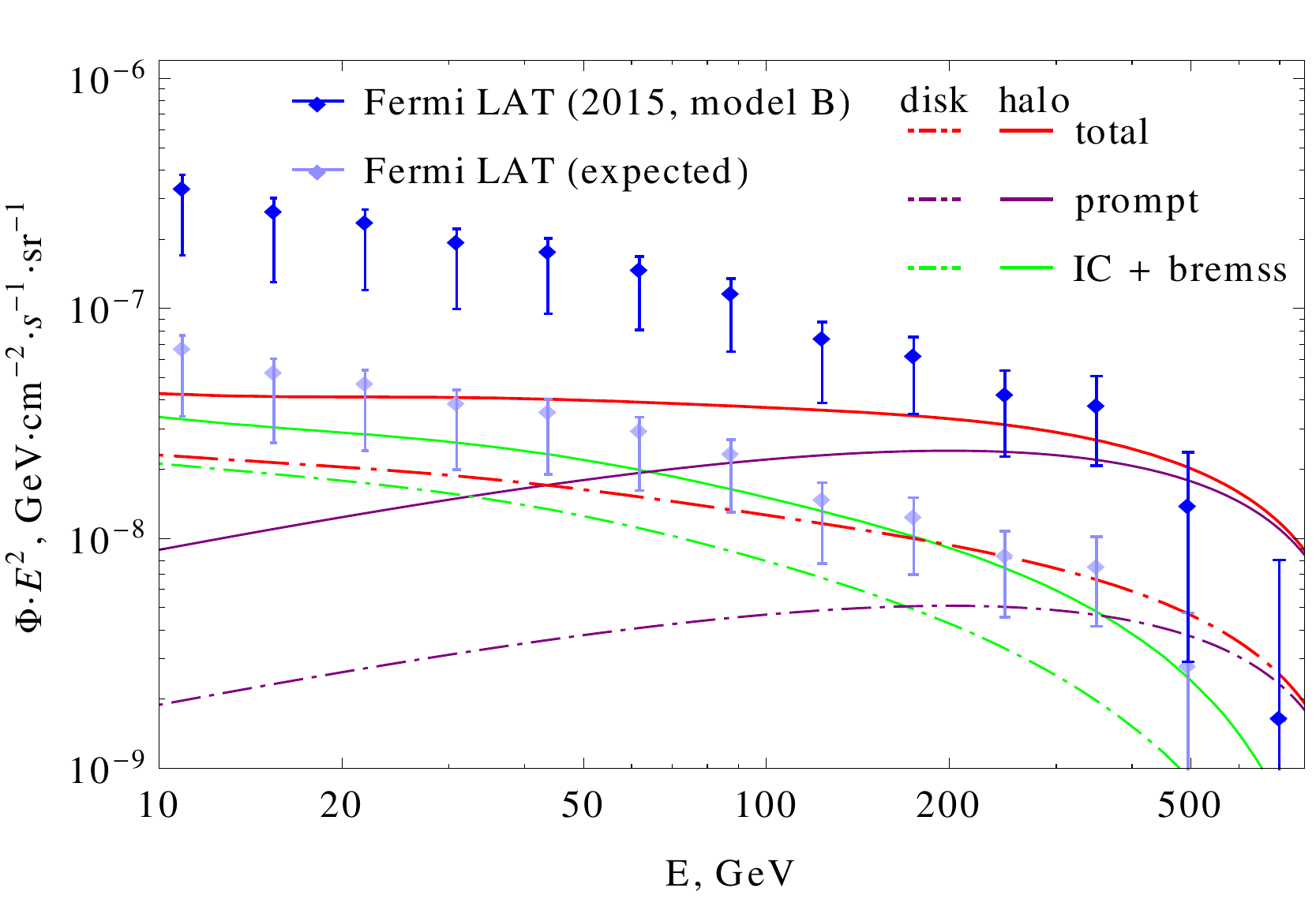} b)}
	\end{minipage}
	\caption{ The fluxes of CR corresponding to the ``minimal case'': the best-fit fraction of positrons (left), %which allows to extract the optimal DM injection spectra (see text), 
		compared to the AMS-02 data; minimal gamma-ray fluxes multiplied by $E^2$, calculated in the halo case (solid curves) and in the DD case (dot-dashed curves), compared to the existing Fermi-LAT IGRB limit (model B, blue dots) and to the expected limit (pale blue dots). Different contributions to gamma rays are shown.}
	\label{min}
\end{figure}

%To prove it 
For the DM injection positron spectrum we intentionally adopt some analytical function %(our choice was not dictated by any physical reasons)
\begin{equation}
%\frac{dN_e}{dE_0}=C_1 E_0^{-C_2}.
f_e(E_0)=\text{const} \cdot E_0^{-1.5} \; (0 \leq E_0 \leq 1\ {\rm TeV}),
\label{minspec}
\end{equation}
which %in the long run (after the corresponding GALPROP calculation) 
allows %us 
to fit the AMS-02 cosmic positron data very well (see Fig. \ref{min} a). Just as in the analysis above, we calculate ICS photons\footnote{Though the flux of ICS radiation depends on the model of CR propagation, the set of parameters we use results in the smallest possible contribution to gamma rays.} and Bremsstrahlung using GALPROP. 

%The gamma-ray flux, corresponding to the minimal case, is shown in Fig.
As one can see in Fig. \ref{min}~b, the total gamma-ray flux in the halo case (red solid line) is very close to the present Fermi-LAT limit at high energies, but does not significantly exceed it. However, taking into account the recent Fermi-LAT statement \cite{DiMauro:2016cbj} that more than $80 \%$ of IGRB can be the product of unresolved astrophysical sources, one would expect the residual isotropic gamma-ray flux to be $\sim 5$ times lower than the existing constraint. Thus, the predicted ``minimal'' gamma-ray flux is found to be in contradiction with the expected IGRB limit when active DM is distributed isotropically in the Galactic halo. In turn, the case of a DD has a total gamma-ray flux (red dash-doted line) that is consistent with this new strict constraint on the isotropic gamma-ray background.

\section{Conclusions and discussion}\label{discus}

We showed that the existence of a DD, mostly populated by active DM, could cure the DM interpretation of the positron anomaly in CR from the overproduction of gamma rays, which, as we have proved, inevitably appears in any DM model that explains this phenomenon and assumes an isotropic distribution of DM. 

The only properties of the DD which can be derived in a model-independent way from our analysis are its thickness and the local emissivity of DM in it. Properties such as local DM density 
%(in other words, the mass of dark disk) 
or Galactic luminosity require additional assumptions about the nature of active DM and about its distribution over galactic scales.
The same actually holds for the gamma-ray luminosity of other galaxies, including dwarf satellites, and even the luminosity of the whole Universe. 
However, a simple estimation shows that since the fraction of active DM in the Universe is roughly proportional to the DD thickness the extragalactic part of IGRB is hardly as big as the Galactic one. As for DM annihilations in disk galaxies, given the distribution of disk galaxies \cite{Ravindranath:2004nd} and taking a well-motivated exponential density profile for the DD (see \cite{Read:2008fh}, Eq. 3) one can obtain that their contribution to IGRB is negligible.
%\textbf{This can be confirmed if model \cite{Read:2008fh} is accepted for dark disk density profile. In fact, only disk-like galaxies \cite{}[[RUSLAN, PUT HERE NECCESSARY REFERENCE]] can not give sufficient contribution in IGRB. While DM in intergalactic medium would be more reasonable supposed to contribute only in decaying DM case. However, this contribution is hardly big as compared to that from Galactic dark disk, since it depends on fraction of decaying DM component in Universe which is roughly proportional to dark disk thickness as well as gamma-signal from it.
Furthermore, as we have checked, the same DD density profile evades the gamma-ray constraints from the Galactic plane \cite{NASAgamma-rayData} and from the Galactic center \cite{TheFermi-LAT:2015kwa}.
%\textbf{Applying the same dark disk density profile do not lead to stronger constraints from data on gamma-rays from galactic plane (first of all from Galactic center \cite{TheFermi-LAT:2015kwa}).}

Nevertheless, these conclusions are not universal and depend on DM model assumptions %concerning physical properties of DM particles 
and %corresponding mechanisms of DD formation. %Different ones can be speculated.
the features of DD formation. Here we consider some speculative mechanisms.

%However, we think that the extragalactic gamma-ray flux (as well as the luminosity of dwarf spheroidal galaxies) is likely to be suppressed sufficiently in the case of a dark disk, as can be expected in the following speculative mechanisms of dark disk formation.

One mechanism is based on purely gravitational effects of galactic collisions and mergers, which result in the formation of a co-rotating dark matter disk \cite{Read:2008fh}. This mechanism does not require multi-component DM. However, as we mentioned above, the resulting disk cannot be sufficiently dense, which means that decaying DM is strongly disfavoured in this case. To make this mechanism an effective framework for a positron anomaly explanation one should probably consider long-range self-interacting DM models. Since DM particles in the disk are expected to have a very different velocity distribution from that of DM particles in the halo, such interactions can naturally provide a boost of annihilation rate in the disk due to the Sommerfeld-Sakharov enhancement\footnote{As argued in \cite{Belotsky:2015osa}, the enhancement of annihilation rate could be even stronger.}. Dwarf satellite galaxies are not expected to emit many gamma rays because they are not likely to store a DD within them.

The second mechanism implies multi-component DM models with the dominant faint component distributed isotropically in the halo and the subdominant active component which forms a disk mainly due to its non-trivial dynamics. This active component can be introduced by simply adding an \textit{ad hoc} new form of DM to the ``ordinary'' one. As an example, one may consider self-interacting DM particles able to form dark atoms \cite{Fan:2013yva,Foot:2014uba}
and dissipate energy during collisions, which can lead to disk emergence.

One can also think of another scenario, in which active DM is \textit{ab initio} presented on par with a faint component. Suppose some new particles $a^+$ and $b^-$ with opposite dark charges %and similar masses 
bound together play the role of ``ordinary'' DM. If the Universe is asymmetric towards the abundance of their corresponding antiparticles then the free leftovers of $a^+$ and $b^-$ %binding 
will annihilate with free $\bar{a}$ and $\bar{b}$. Since the motion of these particles is dissipative as well, it might be possible for them to form a disk-like structure. The mechanism under consideration also provides a very diluted concentration of active DM in dwarf galaxies that makes them dimmer in gamma rays. 

The list of mechanisms and models given here does not pretend to be comprehensive and fully developed, but rather gives the examples of theoretical endeavours to build an appropriate model.

\section*{Acknowledgments}

%\begin{acknowledgments}
We are grateful to J.-R. Cudell for sharing his ideas, which influenced the course of this research, and helpful remarks. We thank M. Khlopov, C. Kouvaris and S. Rubin for thoughtful discussions. We would also like to thank I.~Cholis for useful suggestions and discussion of the concordance between our results and the results presented in his paper \cite{Cholis:2010px}. The work by MEPhI group was performed within the framework of Fundamental Interactions and Particle Physics Research Center supported by MEPhI Academic Excellence Project (contract No~02.03.21.0005, 27.08.2013). The work of M.~L. is supported by a FRIA grant (F.N.R.S.). The work of K.~B. and R.~B. is supported by grant of RFBR (No~14-22-03048). The work of A.~K. is supported by the grant of the Russian Science Foundation (project No~15-12-10039). 

%\end{acknowledgments}

\bibliography{References}
\bibliographystyle{JHEP}

\end{document}